\documentclass[12pt]{article}%
\usepackage[nosort]{cite}
\usepackage{graphicx}
\usepackage{multicol}
\usepackage{amsfonts}
\usepackage{amssymb}
\usepackage{amsmath}
\usepackage{heck}
\usepackage{afterpage}
\usepackage{setspace}
\usepackage{verbatim}
\usepackage{color}
\usepackage{longtable}
\usepackage{float}
\usepackage{subcaption}
\usepackage{epsfig}
\usepackage{epstopdf}
\usepackage{adjustbox}
\usepackage{tikz}
\usepackage[margin=1in]{geometry}
\usepackage{titletoc}
\usepackage{hyperref}%
\usepackage{mathrsfs}
\usepackage{cancel}

\providecommand{\U}[1]{\protect\rule{.1in}{.1in}}
\newsavebox{\mysavebox}

\hypersetup{colorlinks,citecolor=black,filecolor=black,linkcolor=black,urlcolor=black,pdftex}
\usetikzlibrary{decorations.markings}

\numberwithin{equation}{section}

\newcommand{\be}{\begin{equation}}
\newcommand{\ee}{\end{equation}}

\renewcommand{\d}{\mathrm{d}}

\newcommand{\bh}{\mathrm{bh}}
\newcommand{\thm}{\mathrm{th}}

\usepackage{tikz}
\usetikzlibrary{arrows}
\usetikzlibrary{arrows.meta}
\usetikzlibrary{arrows,decorations.pathmorphing}
\usetikzlibrary{shapes.geometric,calc,arrows, positioning,shapes.misc,decorations.markings}
\tikzset{
  big arrow/.style={
    decoration={markings,mark=at position 1 with {\arrow[scale=2,#1]{>}}},
    postaction={decorate},
    shorten >=0.4pt},
  big arrow/.default=black}

\pgfdeclarelayer{edgelayer}
\pgfdeclarelayer{nodelayer}
\pgfsetlayers{edgelayer,nodelayer,main}
\tikzstyle{none}=[inner sep=0pt]

\tikzstyle{SmallCircle}=[draw, shape=circle, fill=black, inner sep=0pt, minimum size=1mm]
\tikzstyle{BigCircle}=[draw, shape=circle, black, fill=black, inner sep=0pt, minimum size=30pt]
\tikzstyle{BigEllipse}=[draw, shape=ellipse, minimum width=0.5cm, minimum height=1cm, black, fill=black, inner sep=0pt]
\tikzstyle{BigEllipse2}=[draw, shape=ellipse, minimum width=0.37cm, minimum height=0.15cm, black, fill=black, inner sep=0pt]

\tikzstyle{VeryThickLine}=[-, line width=2mm]

\tikzstyle{NodeCross}=[draw, shape=circle, cross out, inner sep=0pt, minimum size=6pt,line width=0.25mm]
\tikzstyle{Circle}=[draw, shape=circle, black, fill=black, inner sep=0pt, minimum size=6pt]
\tikzstyle{circle}=[draw, shape=circle, black, fill=black, inner sep=0pt, minimum size=16pt]
\tikzstyle{Star}=[draw, shape=star, fill=red, star points=8, inner sep=0pt, minimum size=8pt]
\tikzstyle{CircleRed}=[draw, shape=circle, black, fill=red, inner sep=0pt, minimum size=6pt]
\tikzstyle{StarP}=[draw={rgb,255: red,128; green,0; blue,128}, shape=star, fill={rgb,256: red,128; green,0; blue,128}, star points=8, inner sep=0pt, minimum size=12pt]
\tikzstyle{ShadedCircRed}=[draw=red, shape=circle, fill={rgb, 255: red,255; green,114; blue, 118}, inner sep=0pt, minimum size=80pt, line width=0.5mm, fill opacity=0.2]
\tikzstyle{ShadedCircRed2}=[draw=red, shape=circle, fill={rgb, 255: red,255; green,114; blue, 118}, inner sep=0pt, minimum size=10pt]
\tikzstyle{ShadedCircRed3}=[draw=black, shape=rectangle, fill={rgb, 255: red,255; green,114; blue, 118}, inner sep=0pt, minimum size=113pt, line width=0.25mm]
\tikzstyle{ShadedCirc}=[draw=red, shape=circle, fill=white, inner sep=0pt, minimum size=45pt,  fill opacity=1.0,  line width=0.5mm]
\tikzstyle{CircleBlue}=[draw, shape=circle, fill=blue, inner sep=0pt, minimum size=6pt]
\tikzstyle{BigCirclePurple}=[draw, shape=circle, fill={rgb,255: red,191; green,0; blue,191}, inner sep=0pt, minimum size=9pt]
\tikzstyle{CirclePurple}=[draw, shape=circle, fill={rgb,255: red,191; green,0; blue,191}, inner sep=0pt, minimum size=8pt]
\tikzstyle{EmptyCircle}=[draw, shape=circle, inner sep=0pt, minimum size=4pt]
\tikzstyle{GreenCircle}=[draw, shape=circle,  fill={rgb,255: red,80; green,200; blue,120}, inner sep=0pt, minimum size=8pt]
\tikzstyle{BrownCircle}=[draw, shape=circle,  fill={rgb,255: red,210; green,105; blue,30}, inner sep=0pt, minimum size=8pt]
\tikzstyle{CirclePurpleSmall}=[draw, shape=circle, fill={rgb,255: red,191; green,0; blue,191}, inner sep=0pt, minimum size=4pt]
\tikzstyle{BigCircleGreen}=[draw, shape=circle, fill={rgb,255: red,0; green,191; blue,0}, inner sep=0pt, minimum size=12pt]
\tikzstyle{BigCircleBlue}=[draw, shape=circle, fill={rgb,255: red,0; green,0; blue,191}, inner sep=0pt, minimum size=12pt]
\tikzstyle{BigCircleRed}=[draw, shape=circle, fill={rgb,255: red,191; green,0; blue,0}, inner sep=0pt, minimum size=12pt]
\tikzstyle{BrownCircleSmall}=[draw, shape=circle,  fill={rgb,255: red,210; green,105; blue,30}, inner sep=0pt, minimum size=6pt]
\tikzstyle{SmallCirclePurple}=[draw, shape=circle, fill={rgb,255: red,191; green,0; blue,191}, inner sep=0pt, minimum size=4pt]
\tikzstyle{SmallCircleRed}=[draw, shape=circle, fill={rgb,255: red,191; green,0; blue,0}, inner sep=0pt, minimum size=4pt]
\tikzstyle{SmallCircleGreen}=[draw, shape=circle, fill={rgb,255: red,0; green,191; blue,0}, inner sep=0pt, minimum size=4pt]

\tikzstyle{DashedLine}=[-, densely dashed, line width=0.25mm]
\tikzstyle{DottedLine}=[-, dotted, line width=0.25mm]
\tikzstyle{ThickLine}=[-, line width=0.25mm]
\tikzstyle{ArrowLineRight}=[-, -{Stealth[scale=1.25]}, line width=0.25mm, scale=5]
\tikzstyle{ArrowLineRed}=[-, draw={rgb,255: red,191; green,0; blue,0}, -{Stealth[scale=1.75]}, line width=0.1mm, scale=5]
\tikzstyle{RedLine}=[-, draw={rgb,255: red,191; green,0; blue,0}, fill=none, line width=0.5mm]
\tikzstyle{DashedLineThin}=[-, densely dashed, line width=0.125mm, fill=none, draw=black]
\tikzstyle{DottedRed}=[-, dotted, draw={rgb,255: red,191; green,0; blue,0}, fill=none, line width=0.25mm]
\tikzstyle{DashedRed}=[-, densely dashed, draw={rgb,255: red,191; green,0; blue,0}, fill=none, line width=0.25mm]
\tikzstyle{BlueLine}=[-, draw={rgb,255: red,0; green,0; blue,191}, fill=none, line width=0.5mm]
\tikzstyle{DottedBlueLine}=[-, dotted,draw={rgb,255: red,0; green,0; blue,191}, fill=none, line width=0.5mm]
\tikzstyle{ArrowLineBlue}=[-, draw={rgb,255: red,0; green,0; blue,191}, -{Stealth[scale=1.75]}, line width=0.1mm, scale=5]
\tikzstyle{GreenDoubleArrow}=[<->, draw={rgb,155: red,0; green,255; blue,0},  line width= 0.5mm, scale=5]
\tikzstyle{RedDoubleArrow}=[<->, draw={rgb,255: red,255; green,0; blue,0},  line width= 0.5mm, scale=5]
\tikzstyle{BlueDottedLight}=[-, dotted, draw={rgb,255: red,0; green,0; blue,191}, fill=none, line width=0.3mm]
\tikzstyle{BrownLine}=[-, draw={rgb,255: red,210; green,105; blue,30}, fill=none, line width=0.5mm]
\tikzstyle{DottedRed}=[-, dotted, draw={rgb,255: red,191; green,0; blue,0}, fill=none, dotted, line width=0.5mm]
\tikzstyle{DottedPurple}=[-, dotted, draw={rgb,255: red,191; green,0; blue,191}, fill=none, dotted, line width=0.5mm]
\tikzstyle{BlueDottedLight}=[-, dotted, draw={rgb,255: red,0; green,0; blue,191}, fill=none, line width=0.5mm]
\tikzstyle{ArrowLinePurple}=[-, draw={rgb,255: red,191; green,0; blue,191}, -{Stealth[scale=1.75]}, line width=0.5mm, scale=5]
\tikzstyle{DashedLineGreen}=[-, densely dashed, draw={rgb,255: red,74; green,103; blue,65}, line width=0.25mm]
\tikzstyle{LineGreen}=[-, draw={rgb,255: red, 74; green,200; blue,65}, line width=0.5mm]
\tikzstyle{ArrowLineGreen}=[-, draw={rgb,255: red,0; green,191; blue,0}, -{Stealth[scale=1.75]}, line width=0.5mm, scale=5]
\tikzstyle{GreenLine}=[-, draw={rgb,255: red,0; green,191; blue,0}, fill=none, line width=0.5mm]
\tikzstyle{PurpleLine}=[-, draw={rgb,255: red,191; green,0; blue,191}, fill=none, line width=0.5mm]
\tikzstyle{PPurpleLine}=[-, draw={rgb,255: red,191; green,0; blue,191}, fill=none, line width=2.5mm]
\tikzstyle{DPurpleLine}=[-, dotted, draw={rgb,255: red,191; green,0; blue,191}, fill=none, line width=0.5mm]
\tikzstyle{SBrownLine}=[-, draw={rgb,255: red,191; green,0; blue,191}, fill=none, opacity=0.35, line width=2.5mm]
\tikzstyle{DottedBlue}=[-, dotted, draw={rgb,255: red,0; green,0; blue,191}, fill=none, line width=0.5mm]

\tikzset{snake it/.style={decorate, decoration=snake}}

\usetikzlibrary{decorations.markings}

\usetikzlibrary{hobby}

\tikzset{
dashstar/.style={
 dash pattern=on 5pt off 5pt,
 postaction={
  decorate,
  decoration={
   markings,
   mark=between positions 9pt and 1 step 10pt with {
     \node[color=red] {*};
   }
  }
 }
},
dashstarstar/.style={ 
 dash pattern=on 5pt off 10pt,
 postaction={
   decorate,
   decoration={
     markings,
     mark=between positions 10pt and 1
          step 15pt
           with {
            \node at (-2pt,0pt) {\pgfuseplotmark{star}};
            \node at (2pt,0pt) {\pgfuseplotmark{star}};
           }
   }
 }
}
}
\usepackage{pgfplots}
\pgfplotsset{compat=1.16}

\newcommand{\lb}{\left(}
\newcommand{\rb}{\right)}

\newcommand{\ba}{\begin{aligned}}
\newcommand{\ea}{\end{aligned}}

\begin{document}

\begin{flushright}
    UUITP-10/25
\end{flushright}

\date{March 2025}

\title{On the Holographic Dual of a \\[4mm] Symmetry Operator at Finite Temperature}

\institution{PENN}{\centerline{$^{1}$Department of Physics and Astronomy, University of Pennsylvania, Philadelphia, PA 19104, USA}}
\institution{PENNmath}{\centerline{$^{2}$Department of Mathematics, University of Pennsylvania, Philadelphia, PA 19104, USA}}
\institution{Uppsala}{\centerline{$^{3}$Department of Physics and Astronomy, Uppsala University, Box 516, SE-75120 Uppsala, Sweden}}

\authors{
Jonathan J. Heckman\worksat{\PENN,\PENNmath}\footnote{e-mail: \texttt{jheckman@sas.upenn.edu}},
Max H\"ubner\worksat{\Uppsala}\footnote{e-mail: \texttt{max-elliot.huebner@physics.uu.se}}, and
Chitraang Murdia\worksat{\PENN}\footnote{e-mail: \texttt{murdia@sas.upenn.edu}}
}

\abstract{Topological symmetry operators
of holographic large $N$ CFT$_D$'s are dual to dynamical branes in the gravity dual AdS$_{D+1}$.
We use this correspondence to establish a dictionary between thermal expectation values of
symmetry operators in the Euclidean CFT$_D$ and the evaluation of gravitational saddles in the presence
of a dynamical brane. Expectation values of $0$-form symmetry operators in the CFT$_D$ are then related
to branes wrapped on volume minimizing cycles in the bulk, i.e., the
Euclidean continuation of a black hole horizon. We illustrate with some representative examples, including
gravity in AdS$_3$, duality / triality defects in 4D $\mathcal{N} = 4$ Super Yang-Mills theory, and the dual of R-symmetry operators probing 5D BPS black holes.}

\maketitle

\enlargethispage{\baselineskip}

\setcounter{tocdepth}{2}


\newpage

\section{Introduction}

Symmetries lead to important constraints on many quantum systems.
Recent work points to deep topological structures connected with
global symmetries in quantum field theories \cite{Gaiotto:2014kfa}.
In gravity, one expects symmetries to be gauged or broken. It is thus of interest to study symmetries
in holographic systems such as the AdS/CFT correspondence \cite{Maldacena:1997re}
where global symmetries of the boundary CFT$_{D}$ correspond
to gauge symmetries in the bulk AdS$_{D+1}$. Recent arguments have established both in
top down \cite{Apruzzi:2022rei, GarciaEtxebarria:2022vzq, Heckman:2022muc, Heckman:2022xgu} and bottom up \cite{Heckman:2024oot}
holographic setups that the topological symmetry operators of the boundary theory correspond to dynamical branes in the bulk, namely branes that couple to local metric fluctuations.

This has led to a fruitful line of development in two directions. On the one hand, one can
start from a known brane in the bulk, thus producing a symmetry operator and its accompanying
fusion rules in the boundary theory. On the other hand, given a known symmetry of the boundary theory,
holographic considerations \textit{predict} the existence of a corresponding dynamical brane.

In this regard, a fundamental quantity of interest is the overall tension of these bulk branes, i.e., their coupling to gravity in a probe approximation.
More broadly, of course, these branes can, in principle, source non-trivial backreaction of the bulk spacetime geometry.

Our aim in this note will be to track the effects of these branes in the bulk by showing how the on-shell gravitational action is deformed by the presence of this brane.
In the boundary CFT, we propose that this corresponds to calculating the expectation value of a symmetry operator wrapped on a non-trivial cycle.

To establish this correspondence, it is helpful to recall some basic entries in the AdS/CFT dictionary. Recall that the gravitational path integral with prescribed boundary conditions
$\varphi_{\mathrm{bulk}} \sim J_{\mathrm{CFT}}$ corresponds to a choice
of background sources \cite{Gubser:1998bc}:
\begin{equation}
    Z_{\mathrm{grav}}[J] = Z_{\mathrm{CFT}}[J].
\end{equation}

This is especially natural in the case of symmetries, where the presence of a source in the CFT simply corresponds to switching on a background chemical potential.
In the bulk, we simply demand a particular boundary condition for our bulk fields.

To keep things concrete, we primarily focus on $D$-dimensional CFTs
on the manifold $M_D = S^1 \times S^{D-1}$. We shall be interested in
wrapping a zero-form symmetry operator $\mathcal{U}$ on the $S^{D-1}$
factor of $M_D$ and evaluating the deformed partition function:
\begin{equation}
    Z_{\mathrm{CFT}}(\beta , \mathcal{U}) = \mathrm{Tr}(e^{-\beta \widehat{H}} \mathcal{U}),
\end{equation}
in the obvious notation.
For invertible symmetries, there is a natural interpretation of operator insertions such as
$\mathcal{U} = \mathrm{exp}(i \mu \widehat{Q})$ as switching on a complexified chemical potential for a charge operator $\widehat{Q}$.
In the bulk, we interpret this as evaluating the on-shell action with prescribed boundary conditions in the presence of a dynamical brane which we denote by $\widetilde{\mathcal{U}}$.

Our plan in the following sections will be to set up this correspondence in more detail.
In particular, given a known dynamical brane in the bulk, we can determine the thermal expectation value for symmetry operators in the boundary theory.
Conversely, knowing a thermal expectation value allows us to determine the backreaction of the dynamical brane on the bulk.
In the limit where the backreaction is small (i.e., in the probe approximation), we can use this to extract the tension of the brane.

\section{Free Energies}

We consider a large $N$ holographic CFT$_D$ on the background $S^1 \times S^{D-1}$ in which we wrap a
zero-form symmetry operator on $S^{D-1}$ and place it at a particular point on the Euclidean circle.
We shall focus on the case of anti-periodic boundary conditions for our fermions, but it is also of interest to consider the case of periodic boundary conditions.

To frame the discussion to follow, consider a Lorentzian signature theory on $\mathbb{R} \times S^{D-1}$ with Hamiltonian density $\mathcal{H}$ and an invertible symmetry with corresponding charge density operator $\mathcal{Q}$.
Integrating both densities over the $S^{D-1}$ results in operators $\widehat{H}$ and $\widehat{Q}$. We can then label states $\vert E,Q,d\rangle$ by their energy $E$, charge $Q$, and an implicit degeneracy factor $d$. A natural quantity to evaluate is the partition function:
\begin{equation}
    \mathrm{Tr}(e^{-\beta \widehat{H} -\mu \widehat{Q}}) = \underset{E,Q}{\sum} N_{E,Q} e^{- \beta E - \mu Q}.
\end{equation}
This is a bit awkward in the case of generalized symmetries since the corresponding ``chemical potential'' is often purely imaginary, when the symmetry is invertible, and sometimes quantized, when the symmetry is discrete. Moreover,
generalized symmetry operators often support non-trivial topological field theories on the worldvolumes, making a formal analytic continuation in parameters more subtle. Further, ``exponentiating infinitesimal non-invertible symmetries'' is rife with possible ambiguities that must be treated with care.

For all these reasons, it will be more convenient to consider the evaluation of the CFT$_D$ with a zero-form symmetry operator inserted:\footnote{The evaluation of this free energy can, in principle, be scheme dependent. This will not end up being an issue since we evaluate expectation values.}
\begin{equation}\label{eq:ZCFT}
    Z_{\mathrm{CFT}}(\beta, \mathcal{U}) \equiv e^{-\mathcal{F}_{\mathrm{CFT}}(\beta , \mathcal{U})} \equiv \mathrm{Tr}(e^{-\beta \widehat{H}} \mathcal{U}),
\end{equation}
where $\mathcal{F}_{\mathrm{CFT}}$ specifies a free energy.
For invertible symmetries of the form $\mathcal{U} = \exp(i \nu \widehat{Q})$
one can implicitly analytically continue in the chemical potential $\mu = -i \nu$.
For ease of exposition, in what follows, we shall mostly assume the existence of a charge conjugation symmetry such that $\mathrm{Tr}(e^{-\beta \widehat{H}} \mathcal{U}) = \mathrm{Tr}(e^{-\beta \widehat{H}} \mathcal{U}^{\dag})$.\footnote{We expect the on-shell action to have complex phases when topological terms in the bulk evaluate to a non-zero value. This is expected in the case of some non-invertible symmetries, but for the most part, we can bypass this subtlety in the present work.} We also consider charged backgrounds, and explain how our considerations extend in this case as well. More broadly, one should also consider complexified saddle point configurations in the gravity dual, a topic we leave for future work.


We claim that the free energy $\mathcal{F}_{\mathrm{CFT}}$ has a natural interpretation in the Euclidean
bulk as the on-shell action in the presence of a dynamical brane. The main point is that the insertion of a symmetry operator
in the boundary theory has already been interpreted in the ``extrapolate dictionary'' (see \cite{Banks:1998dd}) as obtained by pushing a dynamical brane in the bulk all the way to the boundary, or vice versa.\footnote{For non-abelian symmetries the branes which can fully detach from the boundary are labeled by conjugacy classes, otherwise they attach via a flux tube back to the boundary \cite{Heckman:2024oot}. In this note we assume the brane in question can fully detach from the boundary, leaving other situations for future work.} In the bulk, then, we are interested in evaluating the gravitational path integral in the presence of this Euclidean brane, denoted $\widetilde{\mathcal{U}}$.

To frame the discussion, let us briefly review what happens when no brane is inserted.
In the case of a CFT$_D$ on a thermal circle (i.e., anti-periodic boundary conditions for fermions), there are two dominant saddles that contribute, as given by the thermal AdS solution and the AdS-Schwarzschild solution.
Depending on the size of the thermal circle, one or the other will dominate, and this is specified by the Hawking-Page transition  \cite{Hawking:1982dh,Witten:1998zw}.\footnote{To get a phase transition in the dual CFT one has to scale the size of $S^{D-1}$ to infinite volume \cite{Witten:1998zw}. For the generalization of the Hawking-Page transition to R-charged black holes, see reference \cite{Cvetic:1999ne}.}
The metric for thermal AdS is:
\begin{equation}
\label{eq:th_metric}
    \d s^2_{\thm} = \left(\frac{r^2}{\ell^2} + 1 \right) \d \tau^2 + \left(\frac{r^2}{\ell^2} + 1 \right)^{-1} \d r^2 + r^2 \d\Omega_{D-1}^2,
\end{equation}
where $\ell$ is the AdS radius and $\d\Omega_{D-1}^2$ is the metric for a unit radius $S^{D-1}$.
The radial coordinate $r \in [0, \infty)$ with the conformal boundary at $r \to \infty$.
Note that the Euclidean time $\tau$ can take any period here.

The AdS-Schwarzschild solution is:
\begin{equation}
\label{eq:bh_metric}
    \d s^2_{\bh} = \left(\frac{r^2}{\ell^2} + 1 - \frac{\alpha M}{r^{D-2}} \right) \d \tau^2 + \left(\frac{r^2}{\ell^2} + 1 - \frac{\alpha M}{r^{D-2}} \right)^{-1} \d r^2 + r^2 \d\Omega_{D-1}^2 \, .
\end{equation}
where $M$ is the mass of the black hole and
\begin{equation}
    \alpha = \frac{16 \pi G_N}{(D-1) \text{Vol} \left(\Omega_{D-1} \right)} \, .
\end{equation}
In this coordinate system, the radial coordinate $r \in [r_+, \infty)$ where $r_+$ is the largest solution to the equation
\begin{equation}
    \frac{r^2}{\ell^2} + 1 - \frac{\alpha M}{r^{D-2}} = 0 \, .
\end{equation}
This geometry is smooth and completely regular only if the periodicity of $\tau$ is given by
\begin{equation}
    \beta = \frac{4 \pi \ell^2 r_+}{ D r_+^2 + (D-2) \ell^2} \, .
\end{equation}
By abuse of terminology, we shall refer to $r = r_{+}$ as ``the horizon'' even though we work in Euclidean signature.\footnote{We anticipate there is more to say in Lorentzian signature.}

Evaluating the on-shell actions requires regulating each background with an IR cutoff $r = L$. That being said, the difference
in the actions is finite and given by:
\begin{equation}
    \Delta I \equiv I_{\thm} - I_{\bh} =  \frac{\text{Vol} \left(\Omega_{D-1} \right) r_+^{D-1} \left( r_+^2 - \ell^2 \right)}{4 G_N \left(D r_+^2 + (D - 2) \ell^2\right)} \, .
\end{equation}

\begin{figure}
\centering
\scalebox{0.7}{
\begin{tikzpicture}
	\begin{pgfonlayer}{nodelayer}
		\node [style=none] (0) at (-3.5, 3) {};
		\node [style=none] (1) at (-3.5, 2) {};
		\node [style=none] (3) at (-4.5, 0) {};
		\node [style=none] (4) at (-2.5, 0) {};
		\node [style=none] (5) at (-3.5, 1) {};
		\node [style=none] (6) at (-3.5, -1) {};
		\node [style=none] (7) at (-7.875, 0) {};
		\node [style=NodeCross] (8) at (-6.5, 0) {};
		\node [style=NodeCross] (10) at (-5, 0) {};
		\node [style=NodeCross] (11) at (-3.5, 0) {};
		\node [style=NodeCross] (12) at (-7.875, 0) {};
		\node [style=none] (13) at (-6.5, 2.875) {};
		\node [style=none] (14) at (-6.5, 2.125) {};
		\node [style=none] (15) at (-7.875, 2.775) {};
		\node [style=none] (16) at (-7.875, 2.225) {};
		\node [style=none] (17) at (-5, 2.95) {};
		\node [style=none] (18) at (-5, 2.05) {};
		\node [style=none] (19) at (-6.5, 1.95) {};
		\node [style=none] (20) at (-6.5, 0.25) {};
		\node [style=none] (21) at (-7.875, 0.375) {};
		\node [style=none] (22) at (-7.875, 2) {};
		\node [style=none] (23) at (-5, 1.85) {};
		\node [style=none] (24) at (-5, 0.25) {};
		\node [style=none] (25) at (-3.5, 0.425) {};
		\node [style=none] (26) at (-3.5, 1.75) {};
		\node [style=none] (27) at (5, 3) {};
		\node [style=none] (28) at (5, 2) {};
		\node [style=none] (29) at (4, 0) {};
		\node [style=none] (30) at (6, 0) {};
		\node [style=none] (31) at (5, 1) {};
		\node [style=none] (32) at (5, -1) {};
		\node [style=none] (33) at (0.25, 0) {};
		\node [style=NodeCross] (34) at (1.75, 0) {};
		\node [style=NodeCross] (35) at (3.25, 0) {};
		\node [style=NodeCross] (36) at (5, 0) {};
		\node [style=NodeCross] (37) at (0.25, 0) {};
		\node [style=none] (38) at (1.75, 2.75) {};
		\node [style=none] (39) at (1.75, 2.25) {};
		\node [style=none] (42) at (3.25, 2.875) {};
		\node [style=none] (43) at (3.25, 2.125) {};
		\node [style=none] (44) at (1.75, 2) {};
		\node [style=none] (45) at (1.75, 0.25) {};
		\node [style=none] (46) at (0.25, 0.375) {};
		\node [style=none] (47) at (0.25, 2.125) {};
		\node [style=none] (48) at (3.25, 1.85) {};
		\node [style=none] (49) at (3.25, 0.25) {};
		\node [style=none] (50) at (5, 0.425) {};
		\node [style=none] (51) at (5, 1.75) {};
		\node [style=SmallCircle] (52) at (0.25, 2.5) {};
		\node [style=none] (54) at (5, -5.75) {};
		\node [style=none] (55) at (5, -7.75) {};
		\node [style=none] (58) at (5, -5.75) {};
		\node [style=none] (59) at (5, -7.75) {};
		\node [style=none] (60) at (0.25, -6.75) {};
		\node [style=NodeCross] (61) at (1.75, -6.75) {};
		\node [style=NodeCross] (62) at (3.25, -6.75) {};
		\node [style=NodeCross] (63) at (5, -6.75) {};
		\node [style=NodeCross] (64) at (0.25, -6.75) {};
		\node [style=none] (69) at (1.75, -4.75) {};
		\node [style=none] (70) at (1.75, -6.5) {};
		\node [style=none] (71) at (0.25, -6.375) {};
		\node [style=none] (72) at (0.25, -4.625) {};
		\node [style=none] (73) at (3.25, -4.9) {};
		\node [style=none] (74) at (3.25, -6.5) {};
		\node [style=none] (75) at (5, -6.5) {};
		\node [style=none] (76) at (5, -5) {};
		\node [style=SmallCircle] (77) at (0.25, -4.25) {};
		\node [style=none] (82) at (2.875, -4.25) {};
		\node [style=none] (83) at (3.625, -4.25) {};
		\node [style=none] (84) at (3.25, -3.875) {};
		\node [style=none] (85) at (3.25, -4.625) {};
		\node [style=none] (86) at (1.5, -4.25) {};
		\node [style=none] (87) at (2, -4.25) {};
		\node [style=none] (88) at (1.75, -4) {};
		\node [style=none] (89) at (1.75, -4.5) {};
		\node [style=none] (90) at (13.25, -5.75) {};
		\node [style=none] (91) at (13.25, -7.75) {};
		\node [style=none] (92) at (13.25, -5.75) {};
		\node [style=none] (93) at (13.25, -7.75) {};
		\node [style=none] (94) at (8, -6.75) {};
		\node [style=NodeCross] (95) at (9.75, -6.75) {};
		\node [style=NodeCross] (96) at (11.5, -6.75) {};
		\node [style=NodeCross] (97) at (13.25, -6.75) {};
		\node [style=NodeCross] (98) at (8, -6.75) {};
		\node [style=none] (99) at (9.75, -4.9) {};
		\node [style=none] (100) at (9.75, -6.5) {};
		\node [style=none] (101) at (8, -6.4) {};
		\node [style=none] (102) at (8, -4.9) {};
		\node [style=none] (103) at (11.5, -4.925) {};
		\node [style=none] (104) at (11.5, -6.5) {};
		\node [style=none] (105) at (13.25, -6.5) {};
		\node [style=none] (106) at (13.25, -5) {};
		\node [style=none] (112) at (10.985, -4.25) {};
		\node [style=none] (113) at (12.015, -4.25) {};
		\node [style=none] (114) at (11.5, -3.735) {};
		\node [style=none] (115) at (11.5, -4.765) {};
		\node [style=none] (116) at (9.3, -4.25) {};
		\node [style=none] (117) at (10.2, -4.25) {};
		\node [style=none] (118) at (9.75, -3.8) {};
		\node [style=none] (119) at (9.75, -4.7) {};
		\node [style=none] (120) at (12.725, -4.25) {};
		\node [style=none] (121) at (13.775, -4.25) {};
		\node [style=none] (122) at (13.25, -3.725) {};
		\node [style=none] (123) at (13.25, -4.775) {};
		\node [style=none] (124) at (4.5, -4.25) {};
		\node [style=none] (125) at (5.5, -4.25) {};
		\node [style=none] (126) at (5, -3.75) {};
		\node [style=none] (127) at (5, -4.75) {};
		\node [style=none] (128) at (7.635, -4.25) {};
		\node [style=none] (129) at (8.365, -4.25) {};
		\node [style=none] (130) at (8, -3.885) {};
		\node [style=none] (131) at (8, -4.615) {};
		\node [style=none] (132) at (-5, -2) {$\text{Cone}(S^{D-1})\times S^1$};
		\node [style=none] (133) at (-5, -2.75) {Thermal AdS};
		\node [style=none] (134) at (-2.75, 2.5) {$S^1$};
		\node [style=none] (135) at (-2.45, -1.125) {$S^{D-1}$};
		\node [style=none] (136) at (-1.75, 0) {};
		\node [style=none] (137) at (-0.75, 0) {};
		\node [style=none] (138) at (3.25, -2) {$\text{Cone}(S^{D-1}\times S^1)$};
		\node [style=none] (139) at (6.25, -6.75) {};
		\node [style=none] (140) at (7.25, -6.75) {};
		\node [style=none] (141) at (14.25, -6.75) {$S^1$};
		\node [style=none] (142) at (14.425, -4.25) {$S^{D-1}$};
		\node [style=none] (143) at (11.5, -2) {$\text{Cone}(S^1)\times S^{D-1}$};
		\node [style=none] (144) at (11.5, -2.75) {AdS Black Hole};
		\node [style=none] (145) at (3.25, -2.815) {Interface Geometry};
		\node [style=none] (149) at (8, -3.375) {$S^{D-1}_{\text{min}}$};
		\node [style=none] (150) at (6, -8.75) {};
		\node [style=none] (151) at (-7.875, 3.25) {$S^{1}_{\text{min}}$};
	\end{pgfonlayer}
	\begin{pgfonlayer}{edgelayer}
		\draw [style=ThickLine, bend left=90] (0.center) to (1.center);
		\draw [style=ThickLine, bend right=90] (0.center) to (1.center);
		\draw [style=ThickLine, bend right=90, looseness=0.50] (3.center) to (4.center);
		\draw [style=DashedLine, bend left=90, looseness=0.50] (3.center) to (4.center);
		\draw [style=ThickLine, bend left=45] (3.center) to (5.center);
		\draw [style=ThickLine, bend left=45] (5.center) to (4.center);
		\draw [style=ThickLine, bend left=45] (4.center) to (6.center);
		\draw [style=ThickLine, bend right=315] (6.center) to (3.center);
		\draw [style=ThickLine, in=-180, out=90, looseness=0.50] (7.center) to (5.center);
		\draw [style=ThickLine, in=-90, out=180, looseness=0.50] (6.center) to (7.center);
		\draw [style=ThickLine, bend left=90] (13.center) to (14.center);
		\draw [style=ThickLine, bend right=90] (13.center) to (14.center);
		\draw [style=ThickLine, bend left=90] (15.center) to (16.center);
		\draw [style=ThickLine, bend right=90] (15.center) to (16.center);
		\draw [style=ThickLine, bend left=90] (17.center) to (18.center);
		\draw [style=ThickLine, bend right=90] (17.center) to (18.center);
		\draw [style=DottedLine] (22.center) to (21.center);
		\draw [style=DottedLine] (19.center) to (20.center);
		\draw [style=DottedLine] (23.center) to (24.center);
		\draw [style=DottedLine] (26.center) to (25.center);
		\draw [style=ThickLine, bend left=90] (27.center) to (28.center);
		\draw [style=ThickLine, bend right=90] (27.center) to (28.center);
		\draw [style=ThickLine, bend right=90, looseness=0.50] (29.center) to (30.center);
		\draw [style=DashedLine, bend left=90, looseness=0.50] (29.center) to (30.center);
		\draw [style=ThickLine, bend left=45] (29.center) to (31.center);
		\draw [style=ThickLine, bend left=45] (31.center) to (30.center);
		\draw [style=ThickLine, bend left=45] (30.center) to (32.center);
		\draw [style=ThickLine, bend right=315] (32.center) to (29.center);
		\draw [style=ThickLine, in=-180, out=90, looseness=0.50] (33.center) to (31.center);
		\draw [style=ThickLine, in=-90, out=180, looseness=0.50] (32.center) to (33.center);
		\draw [style=ThickLine, bend left=90] (38.center) to (39.center);
		\draw [style=ThickLine, bend right=90] (38.center) to (39.center);
		\draw [style=ThickLine, bend left=90] (42.center) to (43.center);
		\draw [style=ThickLine, bend right=90] (42.center) to (43.center);
		\draw [style=DottedLine] (47.center) to (46.center);
		\draw [style=DottedLine] (44.center) to (45.center);
		\draw [style=DottedLine] (48.center) to (49.center);
		\draw [style=DottedLine] (51.center) to (50.center);
		\draw [style=ThickLine, bend left=90] (54.center) to (55.center);
		\draw [style=ThickLine, bend right=90] (54.center) to (55.center);
		\draw [style=ThickLine, in=-180, out=90, looseness=0.50] (60.center) to (58.center);
		\draw [style=ThickLine, in=-90, out=180, looseness=0.50] (59.center) to (60.center);
		\draw [style=DottedLine] (72.center) to (71.center);
		\draw [style=DottedLine] (69.center) to (70.center);
		\draw [style=DottedLine] (73.center) to (74.center);
		\draw [style=DottedLine] (76.center) to (75.center);
		\draw [style=ThickLine, bend right=90, looseness=0.50] (82.center) to (83.center);
		\draw [style=DashedLine, bend left=90, looseness=0.50] (82.center) to (83.center);
		\draw [style=ThickLine, bend left=45] (82.center) to (84.center);
		\draw [style=ThickLine, bend left=45] (84.center) to (83.center);
		\draw [style=ThickLine, bend left=45] (83.center) to (85.center);
		\draw [style=ThickLine, bend right=315] (85.center) to (82.center);
		\draw [style=ThickLine, bend right=90, looseness=0.50] (86.center) to (87.center);
		\draw [style=DashedLine, bend left=90, looseness=0.50] (86.center) to (87.center);
		\draw [style=ThickLine, bend left=45] (86.center) to (88.center);
		\draw [style=ThickLine, bend left=45] (88.center) to (87.center);
		\draw [style=ThickLine, bend left=45] (87.center) to (89.center);
		\draw [style=ThickLine, bend right=315] (89.center) to (86.center);
		\draw [style=ThickLine, bend left=90] (90.center) to (91.center);
		\draw [style=ThickLine, bend right=90] (90.center) to (91.center);
		\draw [style=ThickLine, in=-180, out=90, looseness=0.50] (94.center) to (92.center);
		\draw [style=ThickLine, in=-90, out=180, looseness=0.50] (93.center) to (94.center);
		\draw [style=DottedLine] (102.center) to (101.center);
		\draw [style=DottedLine] (99.center) to (100.center);
		\draw [style=DottedLine] (103.center) to (104.center);
		\draw [style=DottedLine] (106.center) to (105.center);
		\draw [style=ThickLine, bend right=90, looseness=0.50] (112.center) to (113.center);
		\draw [style=DashedLine, bend left=90, looseness=0.50] (112.center) to (113.center);
		\draw [style=ThickLine, bend left=45] (112.center) to (114.center);
		\draw [style=ThickLine, bend left=45] (114.center) to (113.center);
		\draw [style=ThickLine, bend left=45] (113.center) to (115.center);
		\draw [style=ThickLine, bend right=315] (115.center) to (112.center);
		\draw [style=ThickLine, bend right=90, looseness=0.50] (116.center) to (117.center);
		\draw [style=DashedLine, bend left=90, looseness=0.50] (116.center) to (117.center);
		\draw [style=ThickLine, bend left=45] (116.center) to (118.center);
		\draw [style=ThickLine, bend left=45] (118.center) to (117.center);
		\draw [style=ThickLine, bend left=45] (117.center) to (119.center);
		\draw [style=ThickLine, bend right=315] (119.center) to (116.center);
		\draw [style=ThickLine, bend right=90, looseness=0.50] (120.center) to (121.center);
		\draw [style=DashedLine, bend left=90, looseness=0.50] (120.center) to (121.center);
		\draw [style=ThickLine, bend left=45] (120.center) to (122.center);
		\draw [style=ThickLine, bend left=45] (122.center) to (121.center);
		\draw [style=ThickLine, bend left=45] (121.center) to (123.center);
		\draw [style=ThickLine, bend right=315] (123.center) to (120.center);
		\draw [style=ThickLine, bend right=90, looseness=0.50] (124.center) to (125.center);
		\draw [style=DashedLine, bend left=90, looseness=0.50] (124.center) to (125.center);
		\draw [style=ThickLine, bend left=45] (124.center) to (126.center);
		\draw [style=ThickLine, bend left=45] (126.center) to (125.center);
		\draw [style=ThickLine, bend left=45] (125.center) to (127.center);
		\draw [style=ThickLine, bend right=315] (127.center) to (124.center);
		\draw [style=ThickLine, bend right=90, looseness=0.50] (128.center) to (129.center);
		\draw [style=DashedLine, bend left=90, looseness=0.50] (128.center) to (129.center);
		\draw [style=ThickLine, bend left=45] (128.center) to (130.center);
		\draw [style=ThickLine, bend left=45] (130.center) to (129.center);
		\draw [style=ThickLine, bend left=45] (129.center) to (131.center);
		\draw [style=ThickLine, bend right=315] (131.center) to (128.center);
		\draw [style=ArrowLineRight] (136.center) to (137.center);
		\draw [style=ArrowLineRight] (139.center) to (140.center);
	\end{pgfonlayer}
\end{tikzpicture}}
\caption{Sketch of the topological transition associated with the Hawking-Page transition. We label the topological models associated with the two phases and their interface. Reading left to right: Transition from thermal AdS, through an interface geometry, to an AdS black hole background. The two central figures are distinctly presented, yet equivalent, and related by an interchange $S^1\leftrightarrow S^{D-1}$. The conformal boundary to all geometries is $M_D=S^1\times S^{D-1}$. Thermal AdS contains a minimal dimension-1 bulk cycle $S^1_{\text{min}}$, in contrast, the AdS black hole background contains a minimal codimension-2 bulk cycle $\gamma=S^{D-1}_{\text{min}}$.}
\label{fig:HarryPotter}
\end{figure}
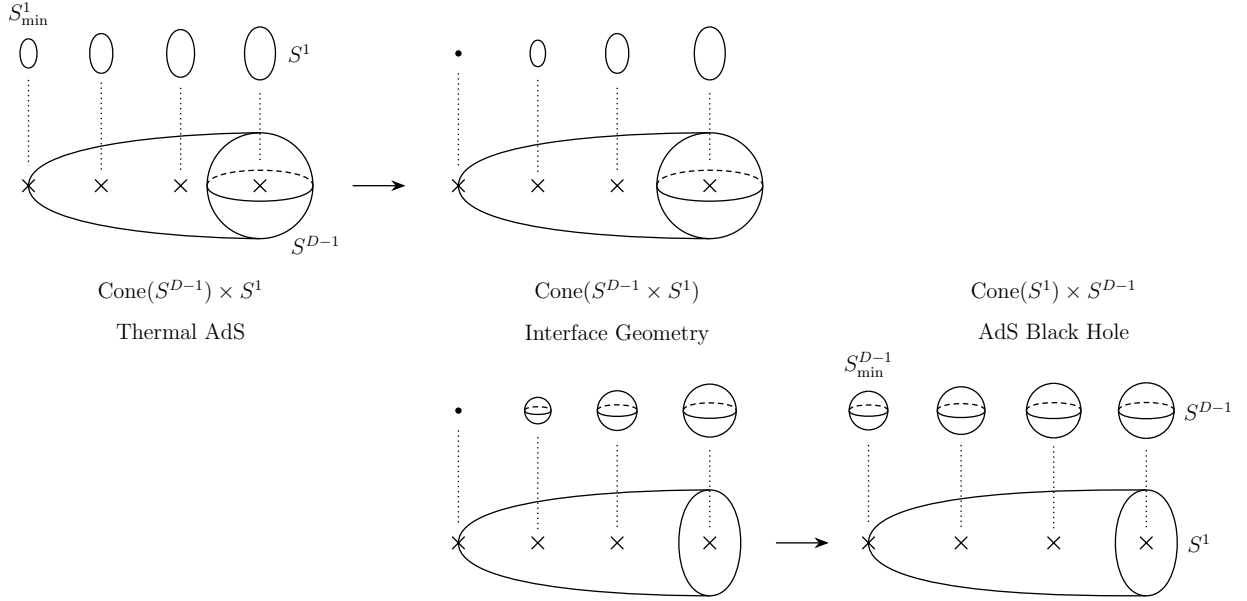

Suppose we now pull our dynamical brane into the bulk. There are two saddle point configurations and thus, two backgrounds probed by the brane.
While the worldvolume theory of the branes is often non-trivial, one expects on general grounds that in a saddle point configuration, the brane will wrap a minimal volume cycle of the bulk. The branes we will consider couple only to the geometric background, and their stable bulk configurations are therefore purely determined by their tension minimizing energy. In the case of thermal AdS, there is no such cycle to wrap; the $S^{D-1}$ can pinch off in the interior. On the other hand, in the AdS-Schwarzschild background, there \textit{is} a minimal volume cycle $\gamma$ given by wrapping the brane on the $S^{D-1}$ at $r = r_{+}$. See figure \ref{fig:HarryPotter}. This minimal cycle has ``area''
\begin{equation}\label{eq:BH}
    A = \text{Vol} \left(\Omega_{D-1} \right) r_+^{D-1} = 4 G_{N} S_{\mathrm{BH}}\,,
\end{equation}
where in the last equality we have used the relation between the area of the outer horizon and the Bekenstein-Hawking (BH) entropy \cite{Bekenstein:1972tm, Bekenstein:1974ax, Hawking:1975vcx}.

Putting this together, we learn that the gravitational path integral with the dynamical brane inserted only affects the black hole saddle:\footnote{More precisely, the value of the action $I_{\thm}$ is unaffected. The boundary 0-form symmetry operator $\mathcal{U}$ is dual in thermal AdS to a flat holonomy along the $S^1$ factor. Further, $I_{\text{th}}$ being unaffected can be anticipated from the CFT dual perspective simply by noting that thermal AdS corresponds to the confined phase, where one expects the degrees of freedom charged under $\mathcal{U}$ to be confined away, suppressing contributions to the leading order as captured by the saddle point approximation. }
\begin{equation}
    Z_{\mathrm{grav}}\big[\widetilde{\mathcal{U}}\big] = \exp(-I_{\thm}) + \exp(-I_{\bh}\big[\widetilde{\mathcal{U}}\big]) \, .
\end{equation}
The thermal expectation of $\mathcal{U}$ is given by the ratio between the two path integrals:
\begin{equation}
\label{eq:Ubeta}
    \langle \mathcal{U} \rangle_{\beta,\mathrm{CFT}} = \frac{Z_{\mathrm{grav}}\big[\widetilde{\mathcal{U}}\big]}{Z_{\mathrm{grav}}} = \frac{\exp(-I_{\thm}) + \exp(-I_{\bh}\big[\widetilde{\mathcal{U}}\big])}{\exp(-I_{\thm}) + \exp(-I_{\bh})}.
\end{equation}

At this level of generality, this is as far as we can go.
However, there are a number of well-motivated approximations
that are natural to consider, leading to more precise estimates.

\subsection{Probe Approximation}

\begin{figure}
\centering
\scalebox{0.8}{
\begin{tikzpicture}
	\begin{pgfonlayer}{nodelayer}
		\node [style=none] (5) at (0, 0.5) {};
		\node [style=none] (6) at (0, -0.5) {};
		\node [style=none] (7) at (3, 2) {};
		\node [style=none] (8) at (3, -2) {};
		\node [style=BigEllipse] (9) at (0, 0) {};
		\node [style=none] (18) at (1.5, 1.25) {};
		\node [style=none] (19) at (1.5, -1.25) {};
		\node [style=none] (20) at (0, -2.5) {$r=r_+$};
		\node [style=none] (21) at (3, -2.5) {$r=\infty$};
		\node [style=BigCircle] (22) at (8.5, 0) {};
		\node [style=none] (23) at (8.5, 2) {};
		\node [style=none] (24) at (8.5, -2) {};
		\node [style=none] (25) at (10.5, 0) {};
		\node [style=none] (26) at (6.5, 0) {};
		\node [style=none] (27) at (8.5, 1) {};
		\node [style=none] (28) at (7.5, 0) {};
		\node [style=none] (29) at (8.5, -1) {};
		\node [style=none] (30) at (9.5, 0) {};
		\node [style=none] (31) at (-1.5, 0) {(i)};
		\node [style=none] (32) at (5.5, 0) {(ii)};
		\node [style=none] (33) at (1.5, 2) {$\widetilde{\mathcal{U}}$};
		\node [style=none] (34) at (9.25, 1.25) {$\widetilde{\mathcal{U}}$};
		\node [style=none] (35) at (8.5, 0) {\color{white} bh};
		\node [style=none] (36) at (0, 1.125) {bh};
	\end{pgfonlayer}
	\begin{pgfonlayer}{edgelayer}
		\draw [style=ThickLine, bend right=15] (7.center) to (8.center);
		\draw [style=ThickLine] (8.center) to (6.center);
		\draw [style=ThickLine] (5.center) to (7.center);
		\draw [style=ThickLine, bend left=15] (7.center) to (8.center);
		\draw [style=BlueLine, bend right=15] (18.center) to (19.center);
		\draw [style=DottedBlue, bend left=15] (18.center) to (19.center);
		\draw [style=ThickLine, bend left=45] (26.center) to (23.center);
		\draw [style=ThickLine, bend left=45] (23.center) to (25.center);
		\draw [style=ThickLine, bend left=45] (25.center) to (24.center);
		\draw [style=ThickLine, bend left=45] (24.center) to (26.center);
		\draw [style=BlueLine, bend right=45] (27.center) to (28.center);
		\draw [style=BlueLine, bend left=45] (27.center) to (30.center);
		\draw [style=BlueLine, bend left=45] (30.center) to (29.center);
		\draw [style=BlueLine, bend right=315] (29.center) to (28.center);
	\end{pgfonlayer}
\end{tikzpicture}
}
\caption{Sketch of the Schwarzschild AdS black hole with Euclidean brane $\widetilde{\mathcal{U}}$ inserted (blue). In (i) we show a profile view, (ii) gives the same geometry head on. In both cases we sketch the $S^{D-1}$ as a circle and do not display $S^1_\beta$, where the brane sits at a point. The brane has finite tension and contracts until it stabilizes, wrapping the black hole horizon $\gamma$ at AdS radius $r=r_+$.}
\label{fig:BH}
\end{figure}
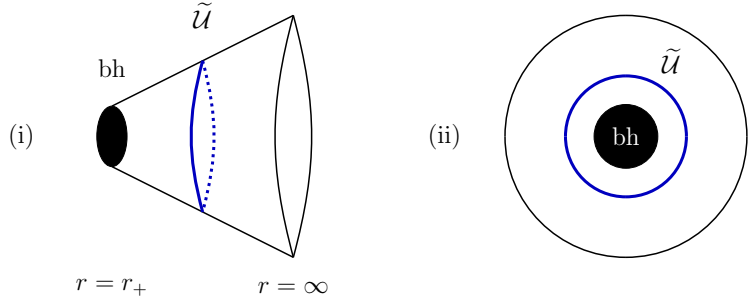

To proceed further, we now assume that the dynamical brane in the bulk produces negligible backreaction.
In this case, the contribution to the on-shell action splits as:
\begin{equation}
\label{eq:split}
    I_{\bh}\big[\widetilde{\mathcal{U}}\big] = I_{\bh} + I_{B}[\gamma] \, ,
\end{equation}
where $I_{B}[\gamma]$ is the on-shell action for the brane wrapped on the minimal volume $S^{D-1}$ (the horizon, see figure \ref{fig:BH}). There can, in principle, be non-trivial dynamics associated with the worldvolume fields of the brane, but all of this can be absorbed into a convention for the overall tension of the brane:\footnote{In many cases $\exp\lb -I_B[\gamma]\rb $ exhibits non-trivial phases since the brane can support topological terms on its worldvolume. Here we focus on the absolute value which quantifies non-topological features of $\widetilde{\mathcal{U}}$.  }
\begin{equation}
\label{eq:tensionapprox}
    \mathrm{Re} (I_{B}[\gamma]) = T_{B} \times A[\gamma] \, ,
\end{equation}
where $T_{B}$ is the tension of the brane in question and $A[\gamma]= \text{Vol} \left(\Omega_{D-1} \right) r_+^{D-1}$.

As we are dealing with a codimension-2 object, it sources a conical deficit angle $\chi$ with $0<\chi<2\pi$.
In the limit where we can isolate the effective tension of the brane, we have:%
\begin{equation}\label{eq:deficitAngle}
T_B = \frac{\chi}{8\pi G_{N}},
\end{equation}
so the on-shell effective action for the brane is:
\begin{equation}
I_{B}[\gamma] = \frac{\chi}{2 \pi} \times S_{\mathrm{BH}}+\dots
\end{equation}
where the ``...'' indicates possible topological terms resulting in imaginary contributions to the on-shell brane action.

Said differently, the change in the on-shell action from including the brane
is proportional to the entropy of the black hole. Let us comment that in
related contexts, it is known that Hubeny-Rangamani-Ryu-Takayanagi (HRRT) surfaces can detach from a boundary
CFT\ and end up wrapping a minimal cycle in the bulk \cite{Hubeny:2007xt}.
It is thus tempting to speculate that the bulk duals of the symmetry operators considered
here are related to HRRT surfaces.

Since we are working in the probe approximation, we implicitly have that $I_{B}[\gamma]$ is a small correction to the bulk gravitational action. As such, we can relate the thermal expectation value and the brane tension. When the black hole saddle dominates, we have:
\begin{equation}
\label{eq:ExpectoPatronum}
    \langle \mathcal{U} \rangle_{\beta , \mathrm{CFT}} = \exp \left( - T_{B} \times A[\gamma] +\dots \right) = \exp\left(- \frac{\chi}{2 \pi} \times S_{\mathrm{BH}} +\dots \right)\,.
\end{equation}
In contrast, when the thermal AdS saddle dominates, the expression \eqref{eq:Ubeta} evaluates to 1.

Beyond this leading order approximation, we can expect additional subleading saddle point configurations to contribute to the gravitational path integral. In the dual CFT, there are also non-perturbative corrections in $N^2 \sim \ell^{D} / G_N$ that must be included.

\section{Illustrative Examples}

Let us give a few illustrative examples. There are of course many further situations which would be exciting to consider, some of which we collect in Appendix \ref{sec:B}.

\subsection{Conical Deficit in AdS$_{3}$}

Many of these considerations become especially tractable in the case of AdS$_3$ / CFT$_2$.
With this in mind, consider the boundary CFT on a rectangular torus (no angular momentum)
described by the coordinates $(\tau,x)$ with periodicities
$\tau \sim \tau + 2 \pi T$ and $x \sim x + 2 \pi R$.\footnote{The inverse temperature is related
to $T$ by $2 \pi T = \beta$. The clash of notation is unfortunate, but standard.}
As in the general case, we have two bulk saddles -- thermal AdS$_3$ and the BTZ black hole.\footnote{One expects additional, but subdominant saddle point configurations. For example, taking a sum over orbits of $SL(2,\mathbb{Z})$ leads to additional contributions, even in pure gravity \cite{Maloney:2007ud}. It would be interesting to study such contributions here, but we defer this to future work.}

\begin{figure}
\centering
\scalebox{0.85}{
\begin{tikzpicture}
	\begin{pgfonlayer}{nodelayer}
		\node [style=none] (12) at (0, -4.5) {(i): Thermal AdS$_3$};
		\node [style=none] (13) at (8, 0.135) {};
		\node [style=none] (14) at (10, 0.135) {};
		\node [style=none] (15) at (12, 0) {};
		\node [style=none] (16) at (6, 0) {};
		\node [style=none] (17) at (8.25,  -0.135) {};
		\node [style=none] (18) at (9.75,  -0.135) {};
		\node [style=none] (19) at (12, -1) {};
		\node [style=none] (20) at (12, 1) {};
		\node [style=none] (24) at (12.75, 0) {$\tau$};
		\node [style=none] (25) at (9, -4.5) {(ii): BTZ Black Hole};
		\node [style=none] (26) at (5, -5) {};
		\node [style=none] (41) at (8, -1) {};
		\node [style=none] (42) at (10, -1) {};
		\node [style=none] (43) at (9, -0.6) {$x$};
		\node [style=none] (44) at (8, -1) {};
		\node [style=none] (45) at (10, -1) {};
		\node [style=none] (46) at (8, 1) {};
		\node [style=none] (47) at (10, 1) {};
		\node [style=none] (48) at (11, 0) {};
		\node [style=none] (49) at (7, 0) {};
		\node [style=none] (50) at (6.5, 0) {};
		\node [style=none] (52) at (0, 3) {};
		\node [style=none] (53) at (0, -3) {};
		\node [style=none] (54) at (0.135, 1) {};
		\node [style=none] (55) at (0.135, -1) {};
		\node [style=none] (56) at (-0.135, -0.75) {};
		\node [style=none] (57) at (-0.135, 0.75) {};
		\node [style=none] (58) at (1, 1) {};
		\node [style=none] (59) at (1, -1) {};
		\node [style=none] (60) at (0, -2) {};
		\node [style=none] (61) at (-1, -1) {};
		\node [style=none] (62) at (-1, 1) {};
		\node [style=none] (63) at (0, 2) {};
		\node [style=none] (64) at (-1, -3) {};
		\node [style=none] (65) at (1, -3) {};
		\node [style=none] (66) at (0, -3.75) {$x$};
		\node [style=none] (67) at (-1, -3) {};
		\node [style=none] (68) at (1, -3) {};
		\node [style=none] (69) at (0, 2.5) {$z=R$};
		\node [style=none] (70) at (9, 0.75) {$z=T$};
		\node [style=none] (71) at (0.75, 0) {$\tau$};
		\node [style=none] (72) at (-3.5, 0) {};
        \node [style=none] (73) at (8, -0.70) {};
		\node [style=none] (74) at (10, -0.65) {};
        \node [style=none] (75) at (6.5, 0) {bh};
	\end{pgfonlayer}
	\begin{pgfonlayer}{edgelayer}
		\draw [style=ThickLine, bend left=90] (16.center) to (15.center);
		\draw [style=ThickLine, bend right=90] (16.center) to (15.center);
		\draw [style=ThickLine, bend right=90, looseness=0.75] (13.center) to (14.center);
		\draw [style=ThickLine, bend left=90, looseness=0.75] (17.center) to (18.center);
		\draw [style=ArrowLineRight, bend right=45] (19.center) to (20.center);
		\draw [style=VeryThickLine, bend right=15, looseness=0.75] (41.center) to (42.center);
		\draw [style=VeryThickLine, bend left=345, looseness=0.75] (47.center) to (46.center);
		\draw [style=VeryThickLine, in=-90, out=15] (45.center) to (48.center);
		\draw [style=VeryThickLine, in=-15, out=90] (48.center) to (47.center);
		\draw [style=VeryThickLine, in=90, out=-165] (46.center) to (49.center);
		\draw [style=VeryThickLine, in=165, out=-90] (49.center) to (44.center);
		\draw [style=ThickLine, bend right=90] (52.center) to (53.center);
		\draw [style=ThickLine, bend left=90] (52.center) to (53.center);
		\draw [style=ThickLine, bend right=90, looseness=0.75] (54.center) to (55.center);
		\draw [style=ThickLine, bend left=90, looseness=0.75] (57.center) to (56.center);
		\draw [style=ArrowLineRight, bend right=15, looseness=0.75] (59.center) to (58.center);
		\draw [style=DashedLine, in=0, out=105] (58.center) to (63.center);
		\draw [style=DashedLine, in=75, out=-180] (63.center) to (62.center);
		\draw [style=DashedLine, bend right=15, looseness=0.75] (62.center) to (61.center);
		\draw [style=DashedLine, in=180, out=-75] (61.center) to (60.center);
		\draw [style=DashedLine, in=-105, out=0] (60.center) to (59.center);
		\draw [style=ArrowLineRight, bend right=45] (64.center) to (65.center);
        \draw [style=ArrowLineRight, bend right=25, looseness=0.65] (73.center) to (74.center);
	\end{pgfonlayer}
\end{tikzpicture}
}
\caption{The metrics of thermal AdS$_3$ and the BTZ black hole parametrize solid tori. We sketch 2-tori with coordinates $x,\tau$, and radial coordinate $z$.  For thermal AdS$_3$ and the BTZ black hole the circles parametrized by $x$ and $\tau$ collapse at $z=R$ and $z=T$, leaving a single circle parametrized by $\tau$ and $x$ respectively (dashed line / thick line).   }
\label{fig:SolidTori}
\end{figure}
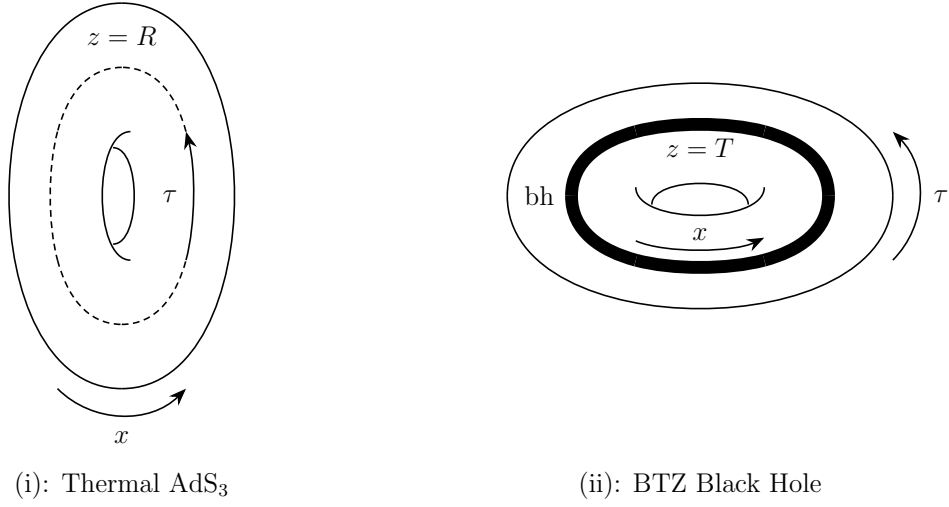

The metric for thermal AdS$_3$ is
\begin{equation}
    \d s^2 = \frac{\ell^2}{z^2} \left( \d \tau^2 + h(z) \d x^2 + \frac{\d z^2}{h(z)} \right) \,\,\, \text{with} \,\,\, h(z) = 1 - \frac{z^2}{R^2} \, ,
\end{equation}
where $ z \in (0, R]$.
For the sake of convenience, we have used a different coordinate choice here, with $z$ being the radial coordinate. The conformal boundary is located at $z = 0$.
In this geometry, the $x$-circle is contractible while the $\tau$-circle is not. See (i) of figure \ref{fig:SolidTori}.

The metric for the BTZ black hole is
\begin{equation}
    \d s^2 = \frac{\ell^2}{z^2} \left( f(z) \d \tau^2 + \d x^2 + \frac{\d z^2}{f(z)} \right) \,\,\, \text{with} \,\,\, f(z) = 1 - \frac{z^2}{T^2} \, ,
\end{equation}
where $ z \in (0, T]$.
In this geometry, the $\tau$-circle is contractible while the $x$-circle is not.  See (ii) of figure \ref{fig:SolidTori}.
After regularization, the Euclidean action of the thermal AdS$_3$ and the BTZ black hole are:
\begin{align}
    I_{\thm} = - \frac{\pi \ell T}{2 G_N R} \, ,
    \qquad\qquad
    I_{\bh}
    &= - \frac{\pi \ell R}{2 G_N T} \, .
\end{align}

Let us now compute the action when we include the backreaction. The Euclidean action of the black hole changes to
\begin{equation}
    I_{\bh}\big[\widetilde{\mathcal{U}}\big] = - \frac{\pi \ell R}{2 G_N T} \left(1 - \frac{8 \pi G_N m}{2 \pi} \right) = - \frac{\pi \ell R}{2 G_N T} + \frac{2 \pi \ell R m}{T} \, .
\end{equation}
Taking the difference with $I_{\bh}$, we observe that we recover the expected tension for a conical defect. By inspection, we can now interpret the parameter $m$ as the tension of the codimension-2 defect, i.e., the mass.

\subsection{Duality / Triality Defects}
\label{sec:DualityTriality}
We now estimate the expectation
value of a duality / triality defect of 4D $\mathcal{N}=4$ Super Yang-Mills
theory.
Bottom-up constructions of these defects were given in \cite{Kaidi:2021xfk, Choi:2022zal}, and a top-down construction was provided in \cite{Heckman:2022xgu, Bashmakov:2022uek}. See also \cite{DelZotto:2024tae}
for related analyses. In what follows, we assume that the complexified gauge coupling $\tau$ has been
tuned, and the global structure of the gauge group has been fixed so that a
duality / triality operation returns us to the same QFT. In terms of top-down
data, this requires a specific choice of the axio-dilaton in the boundary CFT,
as well as a choice of background two-form potentials compatible with this assumption.

The top-down implementation of duality defects arises from a 7-brane with
fixed $\tau_{\text{IIB}}$ wrapped on the $S^{5}$ factor of AdS$_{5}\times
S^{5}$. One can arrange a constant axio-dilaton profile via an appropriate
bound state of $[p,q]$ 7-branes.\footnote{See references \cite{Gaberdiel:1997ud, DeWolfe:1998eu}.}
In particular, the ones of interest for duality / triality defects are the 7-branes associated with specific monodromies /
Kodaira fiber types. We list the relevant 7-branes, including the induced
deficit angles:\footnote{Here we formally continue the pattern, allowing ``$\mathfrak{su}_1$,'' a non-perturbative bound state of 7-branes that corresponds to a non-trivial Kodaira fiber, but with trivial Lie algebra.}
\[%
\begin{tabular}
[c]{|c|c|c|c|c|}\hline
$\tau_{\text{IIB}}$ & Lie Algebra & Kodaira Fiber & 7-branes & $\chi$\\\hline
$\exp(2\pi i/6)$ & $\mathfrak{e}_{8}$ & $II^{\ast}$ & $A^{7}BC^{2}$ &
$10\times\frac{2\pi}{12}$\\\hline
$\exp(2 \pi i/4)$ & $\mathfrak{e}_{7}$ & $III^{\ast}$ & $A^{6}BC^{2}$ &
$9\times\frac{2\pi}{12}$\\\hline
$\exp(2\pi i/6)$ & $\mathfrak{e}_{6}$ & $IV^{\ast}$ & $A^{5}BC^{2}$ &
$8\times\frac{2\pi}{12}$\\\hline
$\exp(2\pi i/6)$ & $\mathfrak{su}_{3}$ & $IV$ & $A^{3}C$ & $4\times\frac{2\pi
}{12}$\\\hline
$\exp(2 \pi i/4)$ & $\mathfrak{su}_{2}$ & $III$ & $A^{2}C$ & $3\times\frac{2\pi
}{12}$\\\hline
$\exp(2\pi i/6)$ & $\mathfrak{su}_{1}$ & $II$ & $AC$ & $2\times\frac{2\pi}%
{12}$\\\hline
\end{tabular}
.
\]

From this, we conclude that the induced deficit angle from a duality defect
leads to an order one contribution to the free energy, so in this sense one might view this as a potentially significant backreaction on the background geometry. On the other hand, we expect that this is still small compared to the contribution from a massive black hole, so we expect a controlled expansion even in this limit. Assuming this approximation holds, we can read off the expectation value of the duality defects, as per equation (\ref{eq:ExpectoPatronum}). As a final comment, observe that there is more than one bound state of 7-branes which produces the same topological duality defect in the dual CFT. This amounts to stacking the minimal TFT by a non-minimal one. In the gravity dual, this additional stacking leads to an increase in the tension.

\subsection{BPS Black Holes in AdS$_{5}$}
\label{ssec:BPSBH}

As a cross-check on the general set of ideas presented here, we now turn to
the case of black holes in 5D $\mathcal{N}=8$ gauged supergravity\footnote{Namely, 32 real supercharges. We follow the conventions of the R-charged black hole literature.} in AdS$_{5}$,
as obtained from consistent truncation of gravity on AdS$_{5}\times S^{5}$.
See e.g. \cite{Gutowski:2004ez, Gutowski:2004yv, Chong:2005da, Kunduri:2006ek,
Hosseini:2017mds} for foundational work on this class of solutions. These are
characterized by three conserved R-charges associated with the angular momenta
on the $S^{5}$, as well as two conserved angular momenta in AdS$_{5}$. We shall be interested in BPS solutions, namely they are extremal (i.e., zero temperature) as well as supersymmetric. This requires all five charges to be switched on. An additional comment is that in Euclidean signature, the condition of supersymmetry requires a Killing spinor which has anti-periodic boundary conditions on the asymptotic boundary $S^1 \times S^3$ \cite{Cabo-Bizet:2018ehj}. Since the ``thermal circle'' at infinity can still be of finite size, we shall, by abuse of terminology, still refer to this as a finite temperature calculation in the dual CFT.

An important recent development is that the black hole entropy can be reproduced
from a corresponding superconformal index for the CFT\ on $\mathbb{R}\times
S^{3}$, provided we work with complexified chemical potentials
\cite{Cabo-Bizet:2018ehj, Choi:2018hmj, Benini:2018ywd, Zaffaroni:2019dhb} (see also \cite{Cabo-Bizet:2019osg, Cabo-Bizet:2020nkr, Benini:2020gjh}). In
Euclidean signature, this involves the index for $\mathcal{N}=4$ Super
Yang-Mills theory with gauge group $SU(N)$ on a spatial $S^{3}$:%
\begin{equation}
    \mathcal{I}(p,q,y_{1},y_{2})=\text{Tr}\left(  (-1)^{F}e^{-\beta\{\mathcal{Q},\overline{\mathcal{Q}}\}}p^{J_{1}+\frac{1}{2}R}q^{J_{2}+\frac{1}{2}R}y_{1}^{q_{1}}y_{2}^{q_{2}} \right) ,
\end{equation}
where here, $\mathcal{Q}$ refers to the supersymmetry generator of a canonical $\mathcal{N} = 1$ subalgebra, and $p,q,y_{1},y_{2}$ are complex fugacities for the various conserved charges. These complex parameters implicitly specify complexified chemical potentials.\footnote{As explained in \cite{Cabo-Bizet:2018ehj}, the anti-periodicity condition for the fermions can be accounted for by a modification of the fugacity $p$.} In the gravity dual, the black hole
entropy follows from the Legendre transform of the on-shell action \cite{Hosseini:2017mds}:\footnote{See also \cite{Ooguri:2004zv, Sen:2005wa, Sen:2008vm, Sen:2009vz}.}
\begin{equation}
    \Gamma =  I-2\pi i\left( \underset{a=1}{\overset{3}{\sum}}Q_{a} \varphi_{a}+\underset{j=1}{\overset{2}{\sum}}J_{j}\omega_{j}\right)  -2\pi i\Lambda\left(  \underset{a=1}{\overset{3}{\sum}}\varphi_{a}-\underset{j=1}{\overset{2}{\sum}}\omega_{j}+1\right) ,
\end{equation}
where we have introduced the chemical potentials $\varphi_{a}$ and $\omega_{j}$ for
the R-charges and angular momenta, $\Lambda$ is a Lagrange multiplier
enforcing a supersymmetry constraint, and $I$ is the on-shell action for a
fixed choice of chemical potentials:%
\begin{equation}
    I=-2\pi i\nu\frac{\varphi_{1}\varphi_{2}\varphi_{3}}{\omega_{1}\omega_{2}} \, ,
    \qquad
    \text{with}
    \qquad
    \nu=\frac{N^{2}}{2}=\frac{\pi \ell^3}{4G_{N}} \, .
\end{equation}
As found in \cite{Cabo-Bizet:2018ehj, Choi:2018hmj, Benini:2018ywd,
Zaffaroni:2019dhb}, the logarithm of the superconformal index is, to leading order in a $1/N$ expansion, simply $I$. A helpful observation is that
\begin{equation}
    \varphi_{a} \frac{\partial\Gamma}{\partial \varphi_{a}}+\omega_{j}\frac{\partial\Gamma}{\partial\omega_{j}}= \Gamma+2\pi i\Lambda \, ,
\end{equation}
so at the critical point, the Lagrange multiplier is proportional to the
entropy:
\begin{equation}
    S_{\text{BH}} = \Gamma_{\text{crit}} = -2\pi i\Lambda_{\text{crit}} \, .
\end{equation}

We now evaluate the expectation value for a $U(1)$ R-charge topological symmetry
operator wrapped on the $S^{3}$, namely
\begin{equation}
    \mathcal{U}_{\alpha} = \exp(2\pi i \Sigma_{a}\alpha_{a}\widehat{Q}_{a}) \, .
\end{equation}
This can be viewed as changing the complexified chemical potential to another value. In the gravity dual, it
suffices to track the change in the on-shell action after this change. Letting
$\varphi_{a}^{\text{crit}}$ denote the critical point used to extremize $\Gamma$, we
have that the change in the on-shell action due to inserting the dynamical brane dual to the symmetry operator is:
\begin{equation}
    I\left(  \varphi_{a}^{\text{crit}}+\delta \varphi_{a}\right)  -I\left(  \varphi_{a}^{\text{crit}}\right)
    = \delta \varphi_{a}\frac{\partial I}{\partial \varphi_{a}} \bigg|_{\text{crit}} \, .
\end{equation}
On the other hand, we also observe that:%
\begin{equation}
    \delta \varphi_{a}\frac{\partial I}{\partial \varphi_{a}} \bigg|_{\text{crit}}
    = \delta \varphi_{a}\left(  \frac{\partial\Gamma}{\partial \varphi_{a}}+2\pi iQ_{a}+2\pi i\Lambda\right)  \bigg|_{\text{crit}} \, .
\end{equation}
Since we are evaluating at a critical point, we have:%
\begin{equation}
    I\left(  \varphi_{a}^{\text{crit}}+\delta \varphi_{a}\right)  -I\left(  \varphi_{a}^{\text{crit}}\right) = \underset{a}{\sum}\delta \varphi_{a}\left( 2\pi i Q_{a}-S_{\text{BH}} \right) .
\end{equation}
To round out the calculation, we observe that on the CFT side of the
calculation, $\delta \varphi_{a} = -\alpha_{a}$, so we get:%
\begin{equation}
    \Delta I=-\underset{a}{\sum}\alpha_{a}\left(  2\pi iQ_{a}-S_{\text{BH}}\right)  ,
\end{equation}
The CFT\ thus yields the expectation value:
\begin{equation}
\label{eq:rot}
    \exp\left(  2\pi i \underset{a}{\sum}\alpha_{a}Q_{a}-\left(  \alpha_{1} + \alpha_{2} + \alpha_{3}\right)  S_{\text{BH}}\right)  = \left\langle \mathcal{U}_{\alpha}\right\rangle_{\mathrm{Index}} \, ,
\end{equation}
where we have introduced the formal ``expectation value'' obtained from the index:
\begin{equation}
    \left\langle \mathcal{U}_{\alpha} \right\rangle_{\mathrm{Index}} = \frac{\text{Tr} \left( (-1)^{F} e^{-\beta\{\mathcal{Q},\overline{\mathcal{Q}}\} } p^{J_{1}+\frac{1}{2}R} q^{J_{2}+\frac{1}{2}R} y_{1}^{q_{1}} y_{2}^{q_{2}} \mathcal{U}_{\alpha} \right)}{\text{Tr} \left(  (-1)^{F} e^{-\beta\{\mathcal{Q},\overline{\mathcal{Q}}\}  } p^{J_{1}+\frac{1}{2}R} q^{J_{2}+\frac{1}{2}R} y_{1}^{q_{1}} y_{2}^{q_{2}} \right)} \, .
\end{equation}

\begin{figure}
\centering
\scalebox{0.8}{
\begin{tikzpicture}
	\begin{pgfonlayer}{nodelayer}
		\node [style=none] (0) at (-1, -1.75) {};
		\node [style=none] (1) at (1, -1.75) {};
		\node [style=none] (2) at (-1, 2) {};
		\node [style=none] (3) at (1, 2) {};
		\node [style=none] (8) at (0.625, 0.25) {};
		\node [style=none] (9) at (-0.625, 0.25) {};
		\node [style=none] (11) at (0, 0) {};
		\node [style=none] (12) at (-0.25, 0.45) {};
		\node [style=none] (13) at (0.25, 0.45) {};
		\node [style=none] (14) at (1.75, -1.75) {$\tau_1$};
		\node [style=none] (15) at (1.75, 0.25) {$\tau_*$};
		\node [style=none] (16) at (1.75, 2) {$\tau_2$};
		\node [style=none] (17) at (0, -2.5) {bh};
		\node [style=none] (18) at (-2.5, 0) {AdS$_5$};
		\node [style=none] (19) at (2.75, 0) {};
		\node [style=none] (20) at (0, -3) {};
		\node [style=none] (21) at (-0.525, 0.925) {$\widetilde{\mathcal{U}}_{\:\!\alpha}$};
		\node [style=none] (22) at (0, -0.75) {};
		\node [style=none] (23) at (0, -0.5) {};
		\node [style=none] (24) at (1, -0.75) {};
		\node [style=none] (25) at (-1, -0.75) {};
		\node [style=none] (26) at (0, -1) {};
		\node [style=none] (29) at (0.7, -0.55) {};
		\node [style=none] (30) at (-0.7, -0.55) {};
		\node [style=none] (31) at (-0.7, -0.95) {};
		\node [style=none] (32) at (0.7, -0.95) {};
		\node [style=none] (33) at (-1, -0.75) {};
		\node [style=none] (34) at (1, -0.75) {};
        \node [style=BigEllipse2] (35) at (0, -1.75) {};
        \node [style=BigEllipse2] (36) at (0, -0.15) {};
        \node [style=BigEllipse2] (37) at (0, 0.15) {};
        \node [style=BigEllipse2] (38) at (0, 2) {};
	\end{pgfonlayer}
	\begin{pgfonlayer}{edgelayer}
		\filldraw[fill=black, draw=gray!50]  (0.2, -1.75) -- (-0.2, -1.75) -- (-0.2, -0.15) --  (0.2, -0.15)-- cycle;
		\filldraw[fill=black, draw=gray!50]  (0.2, 2) -- (-0.2, 2) -- (-0.2, 0.15) --  (0.2, 0.15)-- cycle;
		\draw [style=ThickLine, bend right=90, looseness=0.50] (2.center) to (3.center);
		\draw [style=ThickLine, bend left=90, looseness=0.50] (2.center) to (3.center);
		\draw [style=ThickLine, bend right=90, looseness=0.50] (0.center) to (1.center);
		\draw [style=DottedLine, bend left=90, looseness=0.50] (0.center) to (1.center);
		\draw [style=ThickLine] (3.center) to (1.center);
		\draw [style=ThickLine] (2.center) to (0.center);
		\draw [style=BlueLine, in=-180, out=-90, looseness=0.75] (9.center) to (11.center);
		\draw [style=BlueLine, in=-90, out=0, looseness=0.75] (11.center) to (8.center);
		\draw [style=BlueLine, in=90, out=0, looseness=0.75] (13.center) to (8.center);
		\draw [style=BlueLine, in=90, out=-180, looseness=0.75] (12.center) to (9.center);
	\end{pgfonlayer}
\end{tikzpicture}
}
\caption{We sketch a BPS black hole in AdS$_5$ with global time coordinate $\tau$. The spatial slice at time $\tau_*$ contains a Euclidean brane $\widetilde{\mathcal{U}}_{\alpha}$ linking the world line of the black hole. In particular, the topological worldvolume terms of this brane measure the black hole charges.
}
\label{fig:LinkingBH}
\end{figure}
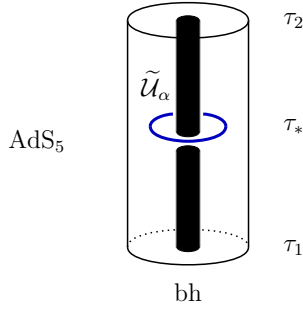

Turning to the gravity dual, this is exactly the answer we should have
expected! The corresponding dynamical brane wraps the outer
horizon at $r=r_{+}$. In particular, in the approximation where backreaction
can be neglected, the brane sources a deficit angle:\footnote{This result
follows from reference \cite{Cvetic:2025kdn}, using results obtained in
reference \cite{Arav:2024exg}. Note that we are working in the limit where $\alpha_{j}$ takes values in $[0,1)$ and is implicitly being treated as small; this accounts for the fact that our expression is not a periodic function of the $\alpha{_j}$.}%
\begin{equation}
    \chi = 2 \pi (\alpha_{1} + \alpha_{2} + \alpha_{3}) \, .
\end{equation}
So, we get the expected contribution to the real part of the on-shell action.
Additionally, the imaginary part follows from the dynamical brane linking with the
charged black hole (viewed as a heavy line operator in the bulk, see figure \ref{fig:LinkingBH}). The black hole carries charges $Q_a$, sourcing a background which is picked up by the brane according to \eqref{eq:rot} whose imaginary part is associated with the usual phase rotation of the symmetry action. All told, we get an exact match!

\section{Conclusions and Future Directions}

In this note, we have provided some further evidence for the correspondence between bulk dynamical branes and boundary theory topological symmetry operators. In particular, we have related the gravity dual on-shell action in the presence of this dynamical brane to the thermal expectation value of the dual symmetry operator in the boundary CFT. This also leads to a number of natural follow-up questions that would be interesting to explore in the future.

An intriguing feature of the bulk calculation is the appearance of a brane wrapping a volume minimizing cycle in the bulk.
This is quite reminiscent of similar structures found in the bulk dual calculation of entanglement entropies in two-sided systems (see e.g., \cite{Hubeny:2007xt}). It would be interesting to see whether there is a deeper reason for this ``phenomenological coincidence.''

There is now some evidence for Hawking-Page transitions in cases where the boundary CFT$_D$ is placed on backgrounds of the form $S^{m} \times S^{n}$ with $n+m = D$ (see e.g., \cite{Aharony:2019vgs}). Wrapping a $p$-form symmetry operator on $S^{m}$ (with $p = D-m-1$) then provides a natural extension of the present case, where now the dynamical brane in the bulk wraps a minimal volume $S^{m}$. Similar considerations likely apply to more general boundary topologies.

The results here have focused on the leading order saddle point configurations in the gravity dual. It would be interesting to systematically include subleading corrections. While this is a challenging problem, in asymptotically AdS$_3$ backgrounds, it should be possible to say more.

Our discussion has centered on Euclidean signature computations. It would be interesting to study the Lorentzian signature case further. For example, the AdS-Schwarzschild geometry has two boundaries. Our current analysis can be viewed as tracking the fate of a single
symmetry operator inserted on just one boundary, and the effect on
an initial state prepared via the Euclidean path integral.\footnote{
Recall that the thermofield double for
a pair of CFTs $\text{CFT}_{L}\otimes \text{CFT}_{R}$ is believed to be dual to the
Lorentzian AdS-Schwarzschild solution \cite{Maldacena:2001kr}. This is given by the
pure state:%
\begin{equation}
    \left\vert \Psi_{\beta}\right\rangle =\frac{1}{\sqrt{Z_{\beta}}} \underset{E=E_{L}=E_{R}}{{\displaystyle\sum}}e^{-\beta E/2}\left\vert E_{L}\right\rangle \otimes\left\vert E_{R} \right\rangle \, .
\end{equation}
We can insert a symmetry operator $\mathcal{U}_{R}$ that acts only on the CFT$_{R}$.
Then, our thermal expectation value calculation amounts to the vev:%
\begin{equation}
    \left\langle \Psi_{\beta}\right\vert \mathcal{U}_{R}\left\vert \Psi_{\beta}\right\rangle =\frac{\text{Tr}(e^{-\beta H}\mathcal{U}_R)}{\text{Tr}(e^{-\beta H})} \, .
\end{equation}

In the gravity dual, the Euclidean signature path integral is preparing a preferred state at some initial time $t=0$.
Including the brane wrapped on the horizon amounts to changing this initial wavefunction.}
It would also be interesting to consider more general multi-partite systems
with symmetry operators inserted (see e.g., \cite{Balasubramanian:2024ysu}).

\section*{Acknowledgements}

We thank C.-M. Chang and H. Zhang for helpful discussions and collaboration at an early stage of this work.
We thank M. Cveti\v{c} for helpful discussions, collaboration on related work, and comments on an earlier draft.
The work of JJH is supported in part by a University Research Foundation grant at the University of Pennsylvania as well as by BSF grant 2022100.
The work of JJH and CM is supported by DOE (HEP) Award DE-SC0013528.  The work of MH is supported by the Marie Skłodowska-Curie Actions under the European Union’s Horizon 2020 research and innovation programme, grant agreement number \#101109804. MH acknowledges support from the
the VR Centre for Geometry and Physics (VR grant No. 2022-06593). The work of CM is also supported by the DOE through QuantISED grant DE-SC0020360.

\appendix

\section{Brief Review of the Hawking-Page Transition}

In this Appendix we briefly review the Hawking-Page transition. See \cite{Hawking:1982dh, Witten:1998zw} for further discussion.\footnote{For a discussion of the Hawking-Page transition for R-charged black holes, see reference \cite{Cvetic:1999ne}.} The on-shell Euclidean action for an AdS$_{D+1}$ solution simplifies to:
\begin{equation}
    I_E =  \frac{D}{8\pi G_N \ell^2} \int \d^{D+1} x \, \sqrt{g} \,
\end{equation}
Thus, for the two geometries given by \eqref{eq:th_metric} and \eqref{eq:bh_metric} respectively, we have:
\begin{align}
    I_{\thm} &= \frac{D}{8\pi G_N \ell^2} \text{Vol} \left(\Omega_{D-1} \right) \beta_{\thm} \frac{L^D}{D} \, , \\
    I_{\bh} &= \frac{D}{8\pi G_N \ell^2} \text{Vol} \left(\Omega_{D-1} \right) \beta_{\bh} \frac{L^D - r_+^D}{D} \, .
\end{align}
Here $L$ is the large-distance cutoff and $\ell$ the AdS radius.
Note that the $\tau$-periodicities are different in the two cases because we need the geometries of the hypersurface at $r = L$ to match.
This is achieved by setting
\begin{equation}
    \beta_{\thm} \sqrt{\frac{L^2}{\ell^2} + 1} = \beta_{\bh} \sqrt{\frac{L^2}{\ell^2} + 1 - \frac{\alpha M}{L^{D-2}}} \, .
\end{equation}
In the $L \to \infty$ limit, the Euclidean actions diverge individually but their difference is finite and given by:
\begin{equation}
    I_{\thm} - I_{\bh} =  \frac{\text{Vol} \left(\Omega_{D-1} \right) r_+^{D-1} \left( r_+^2 - \ell^2 \right)}{4 G_N \left(D r_+^2 + (D - 2) \ell^2\right)} \, .
\end{equation}
Clearly, this difference is positive if $r_+ > \ell$, or equivalently, $\beta < \frac{2 \pi \ell}{D-1}$, and the black hole geometry dominates. Otherwise, if $r_+ < \ell$, the thermal geometry dominates. This is the Hawking-Page phase transition \cite{Hawking:1982dh}.

\section{Further Examples}
\label{sec:B}

In this Appendix we present some further examples involving symmetry operators of the boundary CFT and their counterparts in the gravity dual.

\subsection{D-brane Symmetry Operators}

Another well-understood string construction, complementary to the duality / triality defects constructed from 7-branes and discussed in section \ref{sec:DualityTriality}, are settings in which symmetry operators are realized by wrapped D-branes \cite{Apruzzi:2022rei, GarciaEtxebarria:2022vzq, Heckman:2022muc, Heckman:2022xgu}. The tension $T_p^{(s)}$ of a Dirichlet $p$-brane, in string frame, is given by
\be
T_p^{(s)}=\frac{1}{g_s (2\pi)^p\ell_s^{\:\!p+1}}\,,
\ee
with $g_s$ the string coupling and $\ell_s$ the string length scale. As such, we can readily compute the CFT quantity $\langle \mathcal{U} \rangle_{\beta,\text{CFT}}$ through bulk computation as in \eqref{eq:split} and \eqref{eq:tensionapprox}.

Let us consider concretely the near horizon limit of a stack of D3-branes probing the Calabi-Yau orbifold $\mathbb{C}^3/\mathbb{Z}_{3}$ where the $\mathbb{Z}_3$ acts equally on all complex coordinates by a rotation with $\exp(2\pi i/3)$. The near horizon geometry is AdS$_5\times S^5/\mathbb{Z}_3$. Placing the 4D worldvolume theory on $S^1_\beta \times S^3$ we replace AdS$_5$ with thermal AdS \eqref{eq:th_metric} or the AdS Schwarzschild black hole \eqref{eq:bh_metric} to realize the two saddles dominant for small and large $\beta$.

The CFT is characterized at weak 't Hooft coupling by a 4D quiver gauge theory (see e.g., \cite{Douglas:1996sw, Lawrence:1998ja, Kachru:1998ys}), with a $\mathbb{Z}_3$ zero-form symmetry \cite{Gukov:1998kn}. The symmetry operator to this discrete symmetry is constructed from a D3-brane wrapped on a generator $\sigma_1$ of $H_1(S^5/\mathbb{Z}_3)$ \cite{Heckman:2022xgu}. Therefore, for the AdS Schwarzschild black hole where the $S^5/\mathbb{Z}_3$ does not collapse in the bulk, we also have that $\sigma_1$ remains of finite volume throughout the bulk, namely $\text{Vol}(\sigma_1)=2\pi R/3$ in a convention where the $S^5$ has radius $R$.

Then, making similar approximations as leading up to \eqref{eq:tensionapprox} we have
\be
T_B[\sigma_1]=\frac{2\pi \ell}{3g_s (2\pi)^3\ell_s^{\:\!4}}
\ee
where have set the radius $R=\ell$ the AdS radius. This is compatible, upon making a 5D approximation, with an effective deficit angle $\chi$, following \eqref{eq:deficitAngle}, of
\be
\chi= 8\pi G_N \times \frac{2\pi \ell}{3g_s (2\pi)^3\ell_s^{\:\!4}}\,.
\ee
Overall, noting that length scales relate as $\ell^4= 4\pi N g_s  \ell_s^4$, we compute from \eqref{eq:ExpectoPatronum}
\be\ba
\langle \mathcal{U} \rangle_{\beta,\text{CFT}}&=\exp\lb - \frac{2\pi \ell \text{\;\!Vol}[\gamma]}{3 (2\pi)^3g_s \ell_s^{\:\!4}} \rb=\exp\lb- \frac{N }{ 3\pi \ell^{\:\!3}}\text{Vol}[\gamma]\rb =\exp\lb-\frac{4NG_N}{3\pi \ell^3}S_{\text{BH}} \rb \,, \\[0.4em]
\chi &=\frac{8NG_N}{3\ell^3}
\ea \ee
where $\gamma$ is the minimal volume bulk $S^3$ and
where lastly we used \eqref{eq:BH} and \eqref{eq:ExpectoPatronum}. Next, we rewrite the deficit angle expressing the gravitational constant in other parameters of the problem. For this, recall that the 10D Newton constant is $G_N^{(10)}=8\pi^6 g_s^2\ell_s^8$. The 5D Newton constant is then given by
\be
G_N=\frac{G_N^{(10)}}{\text{\:\!Vol}(S^5/\mathbb{Z}_3)}=\frac{24\pi^6 g_s^2\ell_s^8}{4\pi^3r_+^5}=\frac{3\pi \ell^3}{8N^2} \frac{\ell^5}{r_+^5}\,,
\ee
and therefore, with the dimensionless ratio $c=\ell^5/r_+^5$, we have:
\be
\chi=\frac{c\pi}{N}\,.
\ee

\subsection{Nearly Extremal R-Charged Black Holes}

As another example, we consider nearly extremal zero angular momentum R-charged black hole solutions of 5D\ $\mathcal{N}=8$ gauged supergravity in asymptotic AdS$_{5}$. See e.g., references \cite{Behrndt:1998jd, Behrndt:1998ns, Cvetic:1999ne,
Cvetic:1999rb, Chamblin:1999tk, Myers:2001aq, Buchel:2003re, Gubser:2004xx} for various properties of these solutions.\footnote{See in particular \cite{Cvetic:1999ne} for the construction of R-charged black holes in AdS$_4$ / AdS$_5$ / AdS$_7$ in terms of scaled solutions of rotating M5 / D3 / M2-branes on the internal sphere directions.} We take the asymptotic boundary to be $\mathbb{R}\times S^{3}$. Our discussion follows the summary presented in \cite{Gubser:2004xx}.

The metric is \cite{Behrndt:1998jd}:
\begin{align}
    \d s^{2} &  = -H^{-2/3} f \d t^{2} + H^{1/3} \left( f^{-1} \d r^{2} + r^{2} \d\Omega^{2} \right) \, ,\\
    H &  = \underset{a=1}{\overset{3}{{\displaystyle\prod}}}H_a=\underset{a=1}{\overset{3}{{\displaystyle\prod}}}\left( 1 + \frac{q_{a}}{r^{2}}\right) \, , \\
    f &  = 1-\frac{\mu}{r^{2}}+\frac{r^{2}}{\ell^{2}}H \, ,
    \label{eq:ffunction}
\end{align}
where $\ell$ is the AdS radius, $\mu$ parameterizes the deviation from
extremality and the $q_{a}$ are related to the three internal angular momentum
conserved charges $\widetilde{q}_{a}$ via $\widetilde{q}_{a}=\sqrt{q_{a}%
(q_{a}+\mu)}$. The vector potentials for the three $U(1)$'s and the
corresponding three real scalars are:%
\begin{equation}
    A_{t}^{a} = \frac{\widetilde{q}_{a}}{q_{a}+r^{2}} \, ,
    \qquad
    X^{a}=H_{a}^{-1} H^{1/3} \, .
\end{equation}

Assuming we are near extremality, the Bekenstein-Hawking entropy and inverse
temperature are given by:%
\begin{equation}
S_{\text{BH}} = \frac{A}{4G}\simeq\frac{2\pi^{2}}{4G}\sqrt{q_{1}q_{2}q_{3}%
}\,\,\, \text{and} \,\,\, \beta = \frac{\pi}{\ell}\sqrt{\frac{q_{1}q_{2}q_{3}}{\mu-\mu_{\text{crit}}}},
\end{equation}
where $\mu=\mu_{\text{crit}}$ corresponds to the case where $f(r)$ in line
(\ref{eq:ffunction}) develops a double zero (extremality).

Let us now turn to the dual CFT. Working in the grand canonical ensemble, we
can specify this class of configurations by switching on appropriate chemical
potentials. In this case, the unnormalized density matrix is given by
$\rho\left(  \beta,\varphi_{a}\right)  =\exp(-\beta\widehat{H}-\Sigma
_{a}\varphi_{a}\widehat{Q}_{j})$, where the $\widehat{Q}_{a}$ denote the
$U(1)^{3}$ R-charge operators, and we treat $\beta$ and $\varphi_{a}$ as
fixed. We shall be interested in evaluating the expectation value for a
symmetry operator associated with a $U(1)$ R-charge symmetry operator wrapped
on the $S^{3}$, namely $\mathcal{U}_{\alpha}=\exp(2\pi i\Sigma_{a}\alpha_{a}%
\widehat{Q}_{a})$:%
\begin{equation}
    \left\langle \mathcal{U}_{\alpha}\right\rangle_{\beta , \mathrm{CFT}} =\frac{\text{Tr}\left(  \exp (-\beta\widehat{H}-\Sigma_{a}\varphi_{a}\widehat{Q}_{a})\:\!\mathcal{U}_{\alpha}\right)}{\text{Tr}\left(  \exp(-\beta\widehat{H}-\Sigma_{a}\varphi_{a}\widehat{Q}_{a})\right) } \, .
\end{equation}

In the holographic dual, this amounts to evaluating the on-shell Euclidean
action in the presence of the R-charged black hole. We expect that the
corresponding dynamical brane wraps the outer horizon at $r=r_{+}$. We also
anticipate that since the R-charged black hole non-trivially links with this
symmetry operator, we can also expect a non-trivial complex phase to be
induced. The change in the action is thus:
\begin{equation}
\Delta I=-2\pi i\underset{a}{\sum}\alpha_{a}Q_{a}+\frac{\chi}{2\pi}S_{\text{BH}}=-2\pi
i\underset{a}{\sum}\alpha_{a}Q_{a}+\left(  \alpha_{1}+\alpha_{2}+\alpha_{3}\right)
S_{\text{BH}},
\end{equation}
where in the last equality we used the calculation of the brane tension in
\cite{Cvetic:2025kdn}, using results obtained in \cite{Arav:2024exg}.
Summarizing, we find that the thermal expectation value is:%
\begin{equation}
\exp\left(  2\pi i \underset{a}{\sum}\alpha_{a}Q_{a}-\left(  \alpha_{1}+\alpha_{2}+\alpha
_{3}\right)  S_{\text{BH}}\right)  =\left\langle \mathcal{U}_{\alpha
}\right\rangle_{\beta , \mathrm{CFT}} .
\end{equation}

\newpage

\bibliographystyle{utphys}
\bibliography{FreeEnergyBranes}

\end{document}